\documentclass[ 
    aps,
    pra,
    twocolumn,
    letterpaper, 
    10pt,
    superscriptaddress, 
    showpacs,
    showkeys, 
    notitlepage,
    amsmath, 
    amssymb, 
    floatfix 
]{revtex4-1} 
 
\usepackage{graphicx}
\usepackage{dcolumn}
\usepackage{bm}
\usepackage{color}
\usepackage{amssymb}
\usepackage{amsmath}
\usepackage{hyperref}

\newcommand{\bel}{\begin{equation}}
\newcommand{\eel}{\end{equation}}
\newcommand{\be}{\begin{equation*}}
\newcommand{\ee}{\end{equation*}}
\newcommand{\bal}{\begin{eqnarray}}
\newcommand{\eal}{\end{eqnarray}}
\newcommand{\ba}{\begin{eqnarray*}}
\newcommand{\ea}{\end{eqnarray*}}

\newcommand{\refeq}[1]{Eq.~(\ref{#1})}
\newcommand{\reffig}[1]{Fig.~\ref{#1}}

\newcommand{\ev}[1]{\langle #1 \rangle}
\newcommand{\ket}[1]{| #1 \rangle}
\newcommand{\bra}[1]{\langle #1 |}
\newcommand{\Ev}[1]{\left\langle #1 \right\rangle}

\newcommand{\+}{^\dagger}

\newcommand{\br}{\mathbf{r}}
\newcommand{\bk}{\mathbf{k}}

\newcommand{\PP}{\mathcal{P}}

\newcommand{\eps}{\varepsilon}

\renewcommand{\d}[1]{\!d#1\,}

\begin{document} 

\title{ Quantum network of neutral atom clocks }

\author{P. K\'{o}m\'{a}r}
\affiliation{Physics Department, Harvard University, Cambridge,
MA 02138, USA}

\author{T. Topcu}
\affiliation{Physics Department, Harvard University, Cambridge,
MA 02138, USA}
\affiliation{Department of Physics, University of Nevada, Reno, NV 89557, USA}
\affiliation{ITAMP, Harvard-Smithsonian Center for Astrophysics, Cambridge, MA 02138, USA}

\author{E. M. Kessler}
\affiliation{Physics Department, Harvard University, Cambridge,
MA 02138, USA}
\affiliation{ITAMP, Harvard-Smithsonian Center for Astrophysics, Cambridge, MA 02138, USA}

\author{A. Derevianko}
\affiliation{Physics Department, Harvard University, Cambridge,
MA 02138, USA}
\affiliation{Department of Physics, University of Nevada, Reno, NV 89557, USA}
\affiliation{ITAMP, Harvard-Smithsonian Center for Astrophysics, Cambridge, MA 02138, USA}

\author{V. Vuleti\'{c}}
\affiliation{Department of Physics and Research Laboratory of Electronics,
Massachusetts Institute of Technology, Cambridge, MA 02139, USA}

\author{J. Ye}
\affiliation{JILA, NIST, Department of Physics,  University of Colorado,
Boulder, CO 80309, USA}

\author{M. D. Lukin}
\affiliation{Physics Department, Harvard University, Cambridge,
MA 02138, USA}


\date{\today}

\begin{abstract} 

We propose a protocol for creating a fully entangled GHZ-type state of neutral
atoms in spatially separated optical atomic clocks. In our scheme, local
operations make use of the strong dipole-dipole interaction between Rydberg
excitations, which give rise to fast and reliable quantum operations involving
all atoms in the ensemble.  The necessary entanglement between distant ensembles
is mediated by single-photon quantum channels and collectively enhanced
light-matter couplings.
These techniques can be used to create the recently proposed quantum clock
network based on neutral atom optical clocks. We specifically analyze a possible
realization of this scheme using neutral Yb ensembles.

\end{abstract}


\maketitle

The current record in clock accuracy is held by ytterbium and strontium clocks
\cite{Ludlow2015}, capable of reaching $\sim 10^{-18}$  fractional frequency
stability \cite{Hinkley2013, Bloom2014}. Apart from the enormous amount of
effort and innovation, the unprecedented precision and accuracy were attainable
due to the large number of clock atoms ($10^3-10^4$) \cite{Nicholson2015}.
Super-stable clocks enable evaluation of the systematic frequency shift of 
atomic transitions with less avergaging time, which is important to measure fast
transients, e.g. gravitational waves and passing dark-matter clumps
\cite{Derevianko_Nat_2014}.
In our recent work \cite{Komar2014}, we showed that a quantum network of atomic clocks
can result in substantial boost of the overall precision if multiple
clocks are connected in quantum entanglement. The proposed globally entangled
state, Greenberger-Horne-Zeilinger (GHZ) state, is more sensitive to the global
phase evolution of the clock atoms, thus allows for an improved measurement of
the passage of time. If the GHZ state is set up and interrogated in the optimal
way \cite{Kessler2014, Berry2009}, frequency measurements can asymptotically
reach the Heisenberg limit \cite{Hall2012}, associated with the total number of
atoms in the entire network. 
Significant noise reduction has recently been
demonstrated with spin-squeezed states in a single ensemble of atoms 
\cite{Hosten2016}. Efforts are being
made to make both the non-local \cite{Sangouard2011} and local entanglement
distribution \cite{Sorensen1999, Saffman2010} faster and more reliable. 
Of
particular interest are applications of these ideas to neutral atom clocks.

In this Letter, we show how a non-local  GHZ state can be created across
multiple, spatially separated neutral atom clocks with high fidelity. Our
protocol relies on strong Rydberg blockade for enhancing local atom-atom
interaction, collective excitations for enhancing photon-atom interaction, and
single photon quantum channels for reliable remote
connections. We propose and analyze a realization using neutral Yb ensembles,
suitable for the current atomic clock technology. We predict that thousands of
atoms can be entangled to give an overall stability increase of more than an
order of magnitude, compared to non-entangled clock networks. We emphasize that our
protocol, although presented to be used for a network, can also be applied to a
single ensemble.

\begin{figure}  
\centering
\includegraphics[width=0.45\textwidth]{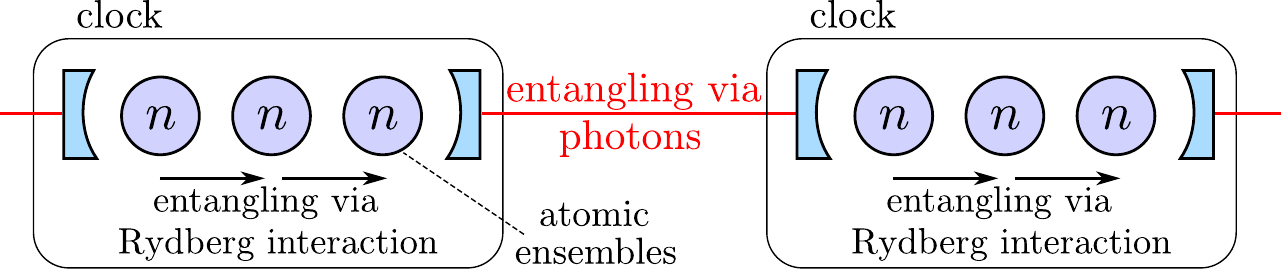}
\caption
[Overview]
{
\label{fig:overview}
(Color online) Schematic of the setup. $K$ clocks, each holding $M$ atomic
ensembles of size $n$ are connected. Atoms within each ensemble get entangled
using long-range interaction between Rydberg atoms, ensembles in the same
clock are entangled  either via Rydberg interactions  or via the cavity
mode, while neighboring clocks are entangled through single-photon quantum channels,
enhanced by optical cavities.
The resulting state is a global GHZ state, $\ket{0}^{\otimes N} + \ket{1}^{\otimes
N}$ of all $N = KMn$ atoms in the network.}
\end{figure}
\begin{figure}
\centering
\includegraphics[width=0.45\textwidth]{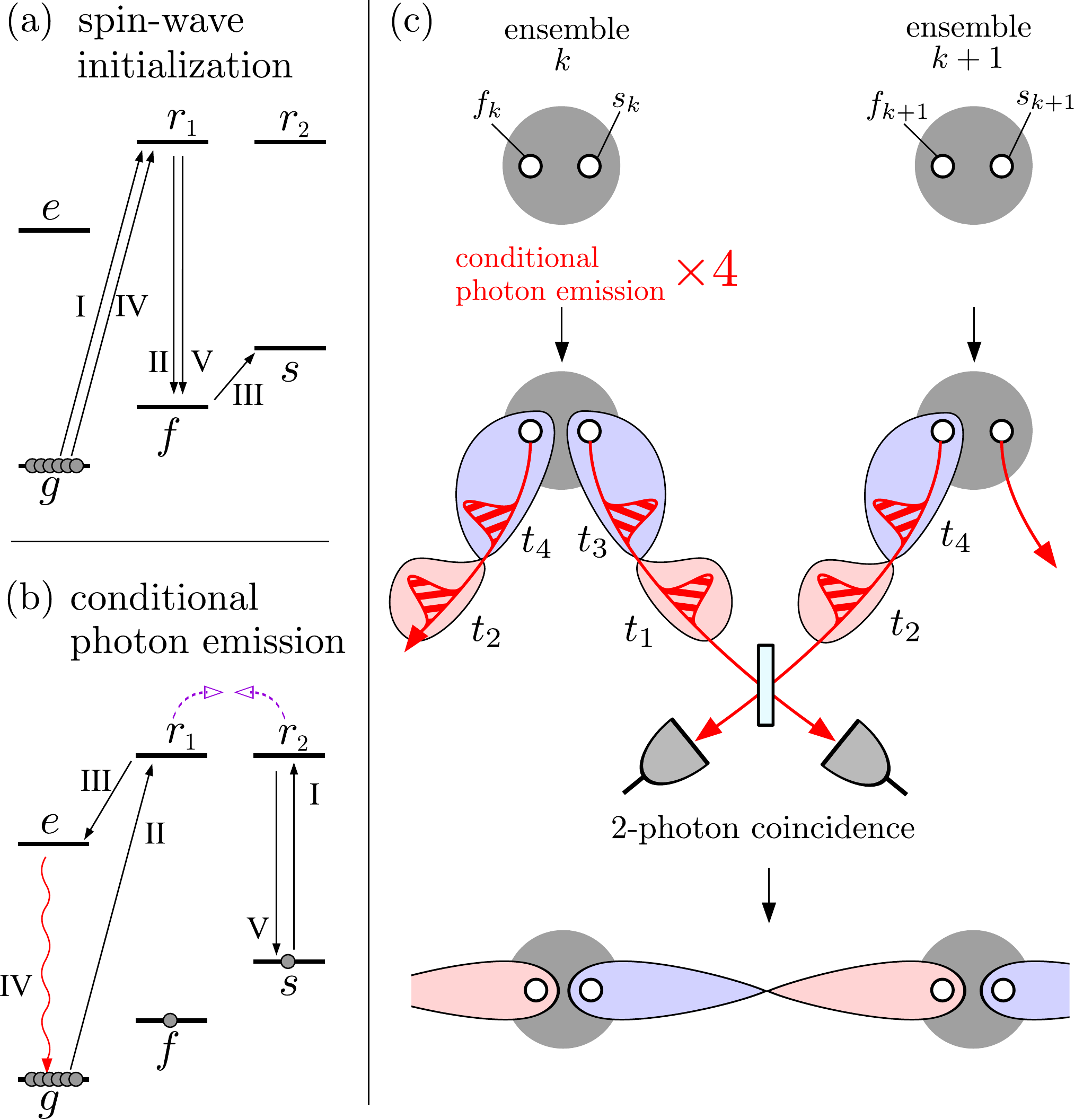}
\caption
[Steps to generate of pairwise entanglement]
{
\label{fig:steps123}
(Color online) Steps to generate pairwise entanglement. (a) Pulse
sequence used to initialize the spin-waves $f$ and $s$ in an ensemble.
(b) Pulse sequence inducing a conditional photon emission, the emitted photon
becomes entangled with the spin state $s$. 
(c) In three steps, neighboring ensembles generate pairwise entanglement between
their collective excitations. First, they induce $0+1$ superpositions of the two
independent spin waves, $f\+$ and $s\+$. Then applying the conditional photon
emission sequence four times, they emit four pulses, containing two
photons total. Each pair of photons is correlated with a unique spin state.
Finally, photons are measured with a linear optics setup, and 2-photon coincidences
indicate the creation of entanglement between neighboring ensembles. (Blue and
red shadings indicate positive and negative correlation between qubits,
respectively.)}
\end{figure}

We describe our protocol for $K$ identical atomic clocks arranged in a sequence,
each connected to its neighbors with optical channels, and each using $Mn$
identical atoms, trapped in a magic-wavelength optical lattice, distributed in
$M$ ensembles, illustrated on \reffig{fig:overview}. 
We use the atomic levels, shown on 
\reffig{fig:steps123}(a) for our protocol:
The two levels of the clock transition, $g, f$, a metastable shelving level $s$, an
excited level $e$, which spontaneously decays to $g$, and two strongly
interacting Rydberg levels, $r_1$ and $r_2$.
We further require transitions between levels, marked with arrows, to be driven
independently.

We imagine preparing all atoms in the ground state $g$, after which
our protocol consists of five subsequent steps.
First, using blockade, we create two independent collective excitations in one
ensemble in each clock, using two separate atomic levels ($f$ and $s$).
Second, each excited ensemble emits single photon pulses that are entangled with
one of these collective excitations.
Third, the photons are sent towards the neighboring atomic clocks, and measured
with a linear optics setup in Bell-basis. Fourth, upon success, each clock
performs a local CNOT operation to connect the two collective excitations. The
result is a set of $K$ entangled collective excitations, one in the first
ensemble of each clock, which serve as "seeds" for a global GHZ state.
In the fifth, and final, step the clocks locally "grow" a GHZ state out of each
seed, extending it to all atoms in the clock, and thus a global GHZ state is
obtained.
In the following, we provide detailed description and analysis of these five
steps, discuss the specific realization in Yb atoms and analyze the most
important sources of imperfections and errors.

Our scheme makes use of the Rydberg blockade, which is a result
of the interaction arising between atoms excited to Rydberg states in an
ensemble. If driven resonantly, the first excited atom blocks the transition of
a second one, thus at most one atom can get coherently excited to the Rydberg
state \cite{Dudin2012, Dudin2010, Ebert2015}, allowing precise quantum control.
Rydberg blockade has been proposed as an efficient tool to realize quantum gates
and perform quantum information processing \cite{Lukin2001, Muller2009,
Saffman2010, Zhao2010, Han2010, Goerz2014}. Efficient control requires the
atoms to reside within the blockade radius of the Rydberg atom.
Different ways of trapping and manipulating Rydberg states are currently  under
investigation both  experimentally \cite{Chen2010, Bariani2012, Firstenberg2013,
Antezza2014, Weber2015} and theoretically \cite{Topcu2013, Beterov2013,
Topcu2014}. 

In the first step, we make use of the Rydberg blockade to create a superposition
of one and zero excitation in both $f$ and $s$ levels, following the approach of
\cite{Lukin2001, Saffman2010,Dudin2012}.
This is done by performing the following sequence of driving pulses:
$[\pi/(2\sqrt{n})]_{g,r1}$, $[\pi]_{f,r1}$, $[\pi]_{f,s}$,
$[(\pi/(2\sqrt{n})]_{g,r1}$, $[\pi]_{f,r1}$, shown in \reffig{fig:steps123}(a),
where $[\phi]_{a,b}$ stands for a pulse with total, single-atom Rabi phase
$\phi$ between level $a$ and $b$.
Starting from the state $\ket{g}^{\otimes n} =: \ket{0}$, this pulse sequence
creates the state
\bel
\label{eq:step1}
(1 + f\+) (1 + s\+) \ket{0} =:
\Big(\ket{0_f} + \ket{1_f}\Big) \Big(\ket{0_s} + \ket{1_s}\Big), 
\eel 
where $f\+$
and $s\+$ are creation operators of the two (approximately) independent spin
wave modes, supported by the two levels $f$ and $s$. 

In the second step, spin-photon entangled states, using the spin wave modes $f$
and $s$, are created, based on an extended version of the scheme described
in \cite{Li2013} and collective enhancement. Each spin-photon entangled state is
created by the pulse sequence shown in \reffig{fig:steps123}(b), involving
$[\pi]_{s,r2}$, $[\pi/\sqrt{n}]_{g,r1}$, $[\pi]_{e,r1}$, $[\pi]_{s,r2}$. With additional pulses
applied before and after this sequence flipping between $0_f\leftrightarrow
1_f$, $0_s\leftrightarrow 1_s$ and swapping $f$ and $s$ waves, and proper
timing, this is repeated four times to produce four time-bin separated light
pulses, which are entangled with the two spin waves,
\bal
	&& \Big(
	\ket{0_f}\ket{t_2} + \ket{1_f}\ket{t_4}
	\Big)
	\Big(
	\ket{0_s}\ket{t_1} + \ket{1_s}\ket{t_3}
	\Big),
\label{eq:step2}
\eal
where $\ket{t_j}\ket{t_k}$ is a two photon state with photons emitted at times
$t_j$ and $t_k$.

In the third step, pairs of time-bin encoded photon pulses from two neighboring
ensembles are detected by interfering the two pulses on a beam splitter and
measuring two-photon coincidences \cite{Duan2001, Honjo2007, Rubenok2013}.
As a result, entangled states between neighboring atomic ensembles, $k$ and
$k+1$, are created \cite{Lukin2003, Shwa2013},
\bel
	\ket{0_s}_k\ket{1_f}_{k+1} \pm \ket{1_s}_k \ket{0_f}_{k+1},
\eel
where the individual kets represent the states of $f$ and $s$ spin waves in the
two ensembles, see \reffig{fig:steps123}(c).

\begin{figure}
\centering
\includegraphics[width=0.45\textwidth]{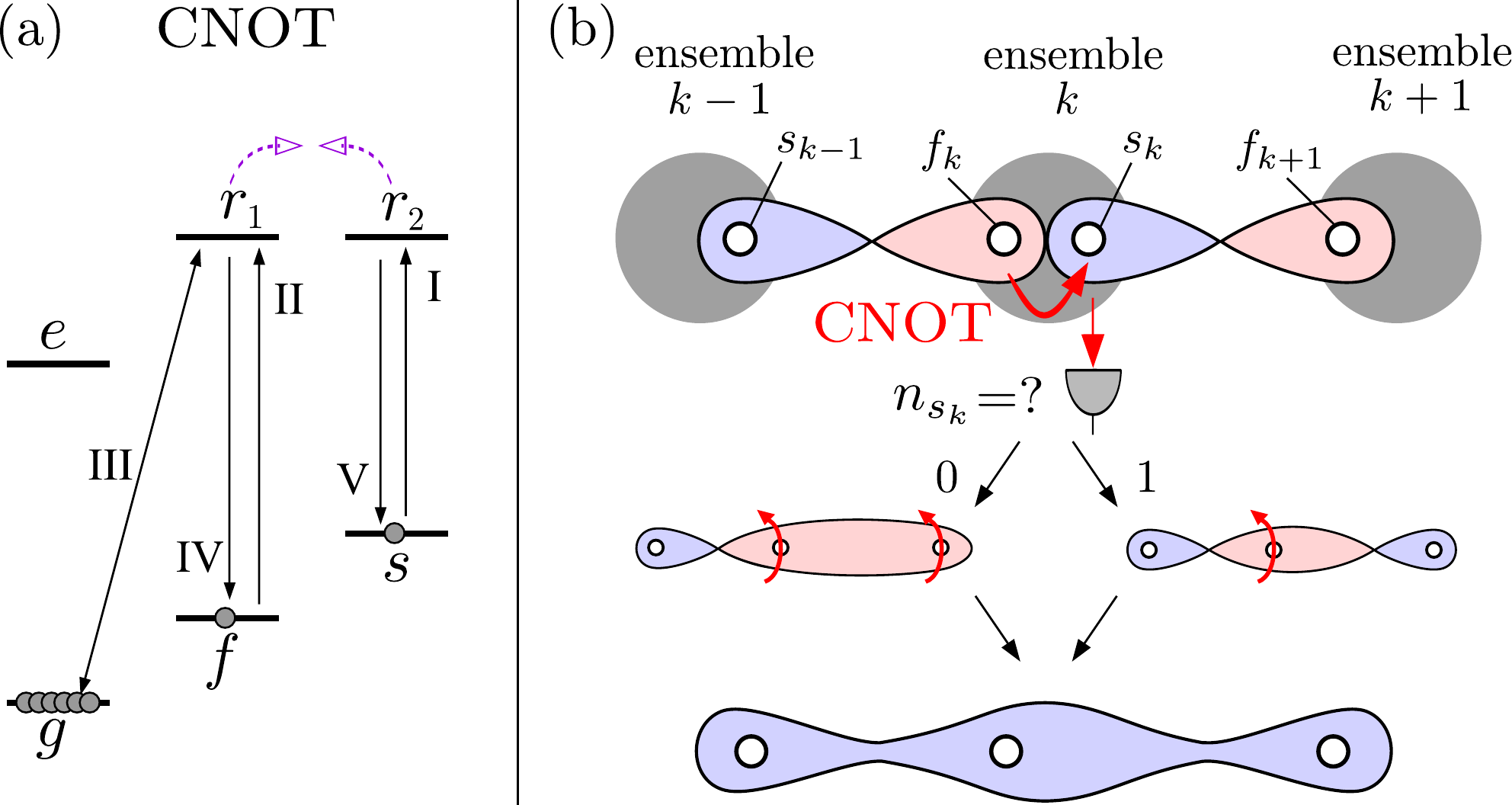}
\caption
[Connecting links into non-local GHZ state]
{
\label{fig:connection} 
(Color online) Connecting links into non-local GHZ state.
(a) CNOT gate between
the two excitations $f$ and $s$: If level $s$ is occupied, then the coherent
(de)excitation of the $f$ level is blocked by the Rydberg blockade between
the $r_1$ and $r_2$ intermediate levels, otherwise it succeeds. 
(b) Connecting two entanglement links. The local CNOT and measurement
operations on ensemble $k$ entangle the two, initially independent, parts of
the system:
$s_{k-1}, f_k$ and $s_k, f_{k+1}$. Depending on the
outcome of the measurement, either only $f_k$, or the entire
right hand side needs to be flipped, in order to arrive to the proper GHZ
state.}
\end{figure}

In the fourth step, the ensembles perform a local CNOT operation on the two
collective degrees of freedom, $f\+$ and $s\+$. This is done with the
following pulse sequence, $[\pi]_{s,r2}$, $[\pi]_{f,r1}$,
$[\pi/\sqrt{n}]_{g,r1}$, $[\pi]_{f,r1}$, $[\pi]_{s,r2}$, shown on
\reffig{fig:connection}(a), which promotes any population in $s$ to $r_2$, which
then blocks the path through $r_1$. The result is a
conditional flip $\ket{0_f} \leftrightarrow \ket{1_f}$, conditioned on having zero $s\+$ excitations. If we perform
$f\leftrightarrow s$ swaps before and after this process, we get a coherent flip
between $\ket{0_f, 0_s} \leftrightarrow \ket{0_f, 1_s}$. 

To understand the resulting state, let us consider two
entangled links, connecting three neighboring ensembles $k-1,k$ and
$k+1$ as shown in \reffig{fig:connection}(b). The corresponding state, before
the fourth step, is
\bel 
	\big(\ket{0,1} + \ket{1,0}\big)_{s_{k-1},f_k}
	\otimes
	\big( \ket{0,1}  + \ket{1,0}  \big)_{s_k,f_{k+1}},
\eel
where $\ket{n_{s_{k-1}}, n_{f_k}}\otimes\ket{n_{s_k},n_{f_{k+1}}}$ indicate the
number of excitations in the modes $s_{k-1}, f_k, s_k, f_{k+1}$ of the three
ensembles.
After the conditional flip of $s_k$ and measurement of $n_{s_k} \rightarrow m
\in \{0,1\}$, the state becomes $\ket{0,1, 1-m} + \ket{1,0,m}$, where the
remaining kets stand for $\ket{n_{s_{k-1}}, n_{f_k}, n_{f_{k+1}}}$.
Depending on the outcome, either only $f_k$ (if $n_{s_k} \rightarrow 1$) or the
entire right hand side (if $n_{s_k} \rightarrow 0$) needs to be flipped in order
to obtain the desired GHZ state, $\bigotimes_k \ket{0_f}_k + \bigotimes_k
\ket{1_f}_k$, of the $f$ excitations of each clock, $k = 1,2,\dots K$.

\begin{figure}
\centering
\includegraphics[width=0.45\textwidth]{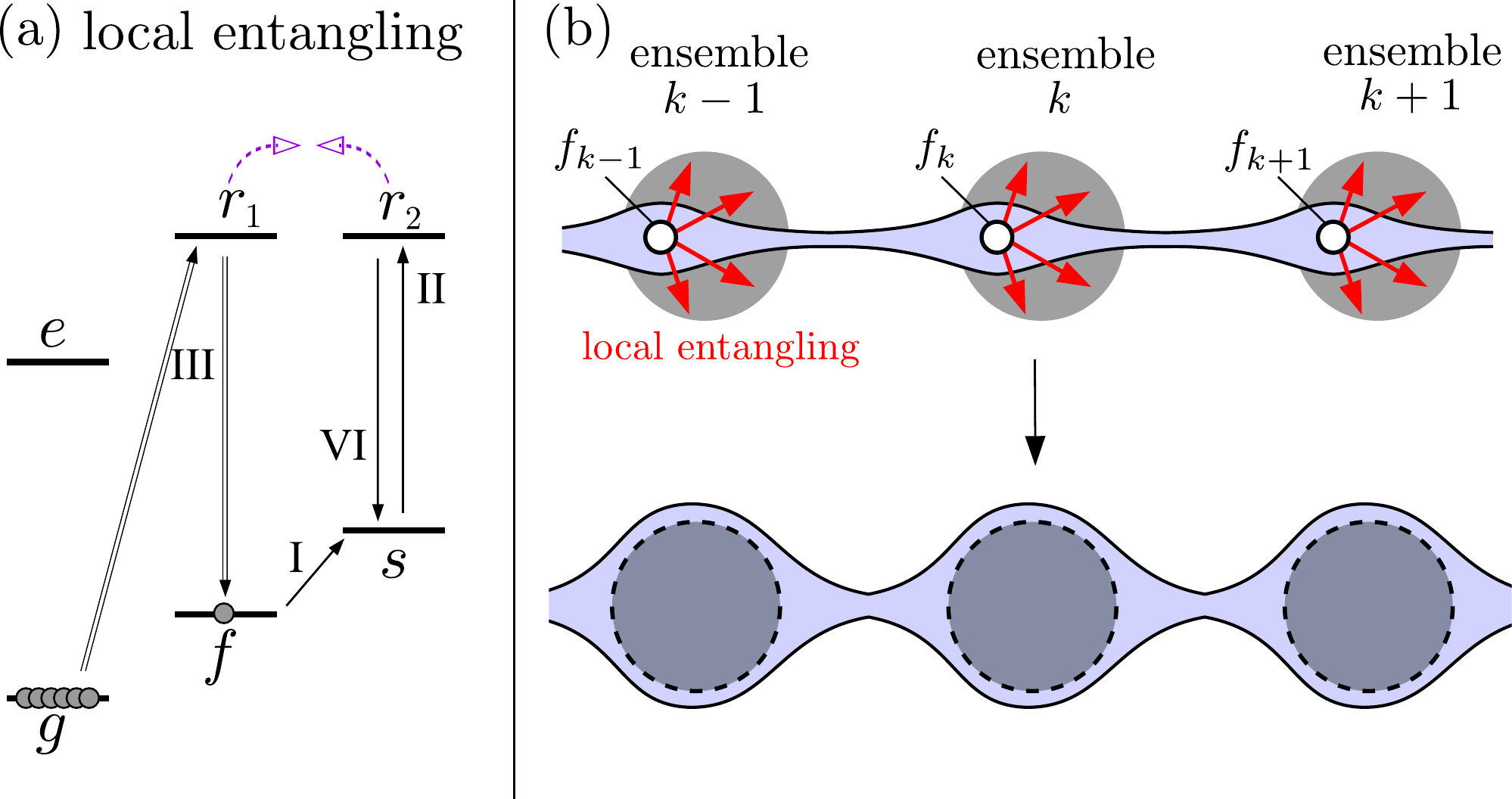}
\caption
[Local GHZ creation]
{  
\label{fig:GHZ}
(Color online) Local GHZ creation.
(a) Conditional, local GHZ state generation: Any excitation in level $s$
prevents the transfer from $g$ to $f$. (b) The local entangling operation
extends the GHZ state from the $f$ spin-wave to all atoms. As a result, every
atom in the network gets entangled.}
\end{figure}

In the fifth step, each clock locally extends the entanglement from its $f$ degree
of freedom to all atoms using a collective Rydberg gate similar to the ones
introduced in Refs.
\cite{Saffman2009, Weimer2010}. In the case when each clock consists of a
single blockaded ensemble, the pulse sequence $[\pi]_{f,s}$, $[\pi/2]_{s,r2}$,
$\left([\pi/\sqrt{n-j+1}]_{g,r1}, [\pi/\sqrt{j}]_{f,r1}\;\text{for}\;
j=1,2,\dots n\right)$, $[\pi]_{s,r2}$, shown in \reffig{fig:GHZ}(a),  does
exactly that.
This sequence transfers the atoms one by one from $g$ to $f$ only if $r_2$ is
unoccupied, and gets blocked otherwise. The result is
\bel
\label{eq:step5}
	\bigotimes_{k=1}^{K}\ket{0_f}_k  + \bigotimes_{k=1}^{K}\ket{1_f}_k
	\;\rightarrow\; \bigotimes_{k=1}^K \ket{f}^{\otimes n} + \bigotimes_{k=1}^K
	s\+\ket{g}^{\otimes n},
\eel
where $\ket{f}$ and $\ket{g}$ denote the state of a single atom.
Finally, we get rid of
the $s$ excitation with a series of pulses that move it back to $g$:
$[\pi]_{f,s}$, $[\pi]_{f,r1}$, $[\pi]_{f,s}$, $[\pi/\sqrt{n}]_{g,r1}$, and end
up with $\ket{f}^{\otimes Kn} + \ket{g}^{\otimes Kn}$, a fully entangled state
of all $N = Kn$ atoms in the network. 

In practice, lattice clocks can employ $n = 10^3-10^4$ atoms each, that can not be manipulated simultaneously 
with high fidelity using Rydberg blockade (see discussion below).  In such a case,  the atoms can be separated into $M \sim 10$ ensembles
within each clock, as shown in Figure 1.  Efficient local entanglement
can be achieved with techniques described in \cite{Sorensen2000} or by using an
individually addressed ``messenger'' atom, that can be moved to the vicinity of
each ensemble to entangle all atoms within each clock using dipole-dipole
interaction. In such a case, the messenger atom can used, first, to extend the
entanglement to all ensembles in each clock, resulting in a state
$\ket{1_f}^{KM} + \ket{0}^{KM}$, after which the procedure shown in
\reffig{fig:GHZ}(a) applied within each ensemble can be used to a fully
entangled state of all $N = K \times Mn$ atoms in the network. (See
Supplementary for details.)

Next, we investigate  the robustness of our  protocol  in  light of realistic
physical imperfections.
We assume that all imperfections decrease the coherence between the two
components of the GHZ state, and therefore the fidelity can be written as
$F = [1 + \exp(-\eps_\text{tot})]/2$, where $\eps_\text{tot}$ is the sum of the
errors.
The errors arising during each non-local connection step
$\eps_\text{non-local}$ and the errors arising during a local GHZ creation
in one clock $\eps_\text{local}$ add up to the total error 
\bel
	\eps_\text{tot} = (K-1)\eps_\text{non-local} + KM\eps_\text{local}.
\eel
This error increases linearly with the total number of atoms in the network,
$N$, and the coefficient, $(\eps_\text{non-local}/M + \eps_\text{local})/n$,
depends on the number of atoms, $n$, within a single atom cloud under blockade.
For a certain optimal local atom number $n_\text{opt}$, the total fidelity is
maximal, i.e. decreases with the slowest rate, as $N$ increases.

\begin{figure}[h]
\centering
\includegraphics[width=0.45\textwidth]{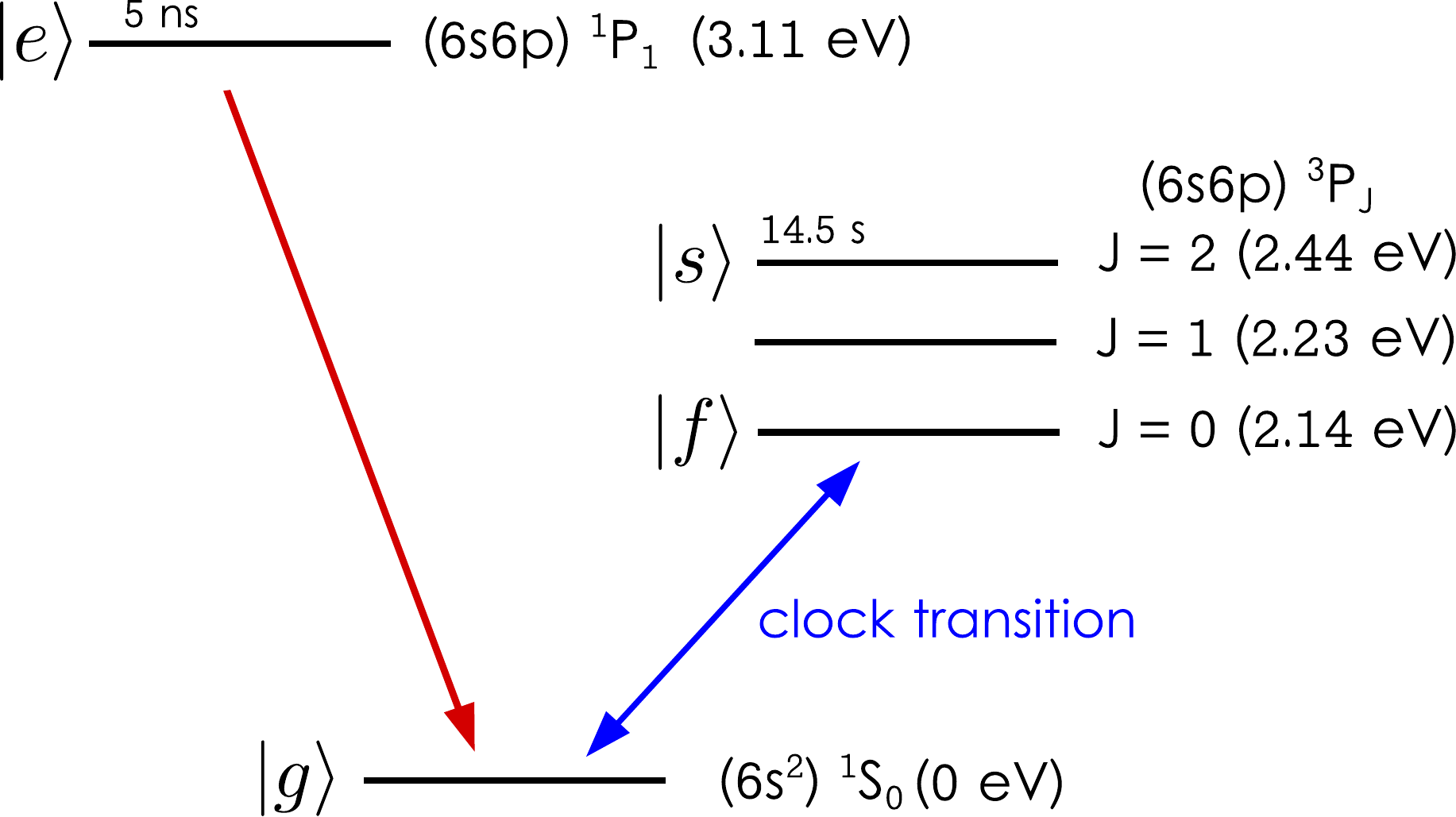}
\caption
[Implementation with Yb levels]
{ 
\label{fig:Yb_levels}
(Color online)
Implementation of our protocol in the lower level of neutral Yb. We assign
the roles of $g$ and $f$ to the clock levels, the role of $s$ to the metastable
$J=2$ level of $6s6p$, and the role of $e$ to the ${}^1P_1$  excited
state, which spontaneously decays to the ground state.}
\end{figure}

To be specific, we focus on a possible implementation of our scheme with
ensembles of neutral ytterbium atoms whose relevant electronic levels are shown on
\reffig{fig:Yb_levels}.
We identify the  following levels of neutral
Yb  relevant for our protocol:
$\ket{g} = \ket{6s^2({}^1\!S_0)}$, 
$\ket{f} = \ket{6s6p({}^3\!P_0)}$, 
$\ket{s} = \ket{6s6p({}^3\!P_2)}$ and 
$\ket{e} = \ket{6s6p({}^1\!P_1)}$, and 
two Rydberg levels 
$\ket{r_1} = \ket{6s\tilde n p_{m=+1}({}^1P_1)}$ and 
$\ket{r_2} = \ket{6s\tilde n s({}^3S_1)}$ 
with the same principle quantum number $\tilde n$. 
Collective enhancement and phase matching of the laser pulses make the emitted
photons leave in a well-defined, narrow solid angle, resulting in high photon
collection efficiency.
Due to the different symmetries of these
states, the coherent coupling can be done via 1-photon transitions for
$r_1\leftrightarrow g$ and $r_2\leftrightarrow s$, and requires 2-photon
transitions for $r_1\leftrightarrow e$ and $r_1\leftrightarrow f$. 
We envision the atoms
being held in position by an optical lattice with period $a =
275.75~\text{nm}$, each potential minimum holding exactly one Yb atom.
(The lattice intensity can be modulated during the Rydberg state excitation
\cite{Tiecke2014}.)  
 
We consider the following errors in our analysis. During
non-local connection, we take into account the finite $r_1$-$r_2$ interaction, which
allows the creation of an $r_1$ excitation with some small probability, even if
$r_2$ is populated, the finite lifetime of the $s$ and $r_2$ levels,
and
the dark-count rate of photo-detectors. For the local GHZ creation step, we account
for the same imperfection of the
$r_1$-$r_2$ blockade as for the non-local entangling step, the finite lifetimes
of the Rydberg levels $r_1$ and $r_2$, and the imperfect self-blockade of the
single excited Rydberg states $r_1$.
(See Supplementary Materials for details.) We estimate the effect of these
errors, and numerically optimize the free parameters: the Rabi frequency
$\Omega$ of the transferring pulses $g\rightarrow r_1$ and $r_1\rightarrow f$,
and the number of local atoms $n$, for principle quantum numbers, $50\leq \tilde
n \leq 150$ of the Rydberg levels, in order to find the minimal error per atom,
$E := \eps_\text{tot}/N$.
 
To illustrate, for Rydberg levels $\tilde n = 120$, we find that the highest
fidelity is reached for $n_\text{opt} \approx 146$, and $\Omega = 10^5\,\gamma$,
where $\gamma \sim 10^3\,\text{s}^{-1}$ is the natural linewidth of the Rydberg
levels, for a clock size of $(Mn)_\text{opt} = 2500$. In this case, the error
per atom is $E_\text{min}= [\eps_\text{tot}/N]_\text{min} = 1.8 \times 10^{-5}$.
Contributions of the different error sources are shown in Table
\ref{table:errors}. We find that the decay of the Rydberg level, and imperfect
blockade cause the majority of imperfections, both arising during the critical
step, local extension of the GHZ state. (See Supplementary Materials for more
details.)
\begin{table}
\centering
\begin{tabular}{|l|c|c|}
\hline
 Errors in 3D ensemble & error per atom & ratio in total\\
\hline
imperfect blockade ($e_1$) 	& $2.6 \times 10^{-6}$ 	& 14\%\\
Rydberg decay ($e_2$) 		& $1.6 \times 10^{-5}$ 	& 86\%\\
self-blockade ($e_3$) 		& $\sim 10^{-11}$ 	& $<0.1$\%\\
$r_2$ decay (non-local) ($e_4$) & $ \sim 10^{-11}$ 	& $<0.1$\%\\
photon detection ($e_5$) 	& $ \sim 10^{-12}$ 		& $<0.1$\%\\
memory error ($e_6$) 		& $ \sim 10^{-8}$ 	& $<0.1$\%\\
photon collection ($e_7$) & $\sim 10^{-8}$ & $<0.1$\%\\
\hline
total error per atom 		& $1.8 \times 10^{-5}$ 	& 100\%\\
\hline
\end{tabular}
\caption
[Error budget]{
\label{table:errors}
The absolute and relative contribution of the different error sources to the
total error per atom $E$, at $\tilde n = 120$, $\Omega = \Omega_\text{opt} = 
10^5\,\gamma$ and $n = n_\text{opt} = 146$, after numerical
optimization, for a 3D lattice.
(See Supplementary Materials for 2D results.)}
\end{table}

With the optimal ensemble size $n_\text{opt}$, determined above, we consider the
total number of entangled atoms $N$. Although having more atoms always results
in improved clock precision, entangling all available atoms is not necessarily
optimal. To see this, we compare the stability of the entangled clock network
and a non-entangled network, and find an optimal entangled atom number
$N_\text{opt}$ by maximizing the stability gain over the non-entangled scheme,
\bel
\label{eq:gain}
	G = \frac{\sigma_\text{non-ent}}{\sigma_\text{ent}/(2F-1)} =
	e^{-EN}\frac{\pi}{8}\sqrt{\frac{N}{\log N}},
\eel 
where $\sigma_\text{ent} = \frac{1}{\omega_0 \tau}\frac{8}{\pi}\frac{\sqrt{\log
N}}{N}$ (from \cite{Komar2014}, assuming perfect fidelity, and that $\tau$ is
smaller than the reduced atomic coherence time $\gamma_\text{at}^{-1}/N$) and
$\sigma_\text{non-ent} = \frac{1}{\omega_0\tau}\frac{1}{\sqrt{N}}$ (for $N$
independent atoms) are the Allan deviations of the two schemes, where $\omega_0$
is the central frequency and $\tau$ is the total available measurement time.
The additional factor of $2F-1 = e^{-EN}$ is due to the reduced Fisher
information of a non-pure GHZ state, where $F$
is the fidelity of the initial state.
(See supplementary materials for details.) For $E = E_\text{min} = 1.8\times
10^{-5}$, \refeq{eq:gain} is maximized with optimal atom number $N_\text{opt}
\approx 1/(2E_\text{min}) \approx 25000$, where $G_\text{max} \sim 12$, and $F
= [1+e^{-N_\text{opt} E_\text{min}}]/2 = 0.82$. The optimal gain  is achieved by
25000 entangled atoms distributed in $K_\text{opt} = N_\text{opt}/(Mn)_\text{opt}
\approx 10$ clocks.


We presented and analyzed a protocol, capable of fully entangling ensembles of
neutral atoms located in different atomic clocks. Local interactions are made
robust by utilizing the strong interaction between Rydberg excitations, and
non-local entanglement creation is made reliable with strong atom-light
coupling, suppressed photon propagation errors and long atomic memory lifetimes.
We showed that our scheme, in particular a realization with neutral Yb
ensembles, is feasible and provides significant gain over non-entangled schemes
even in the light of physical imperfections. Our results provide the first
detailed proposal for a neutral atom clock network that can serve as a first
prototype of the global quantum clock network outlined in \cite{Komar2014}.


We are grateful to Kyle Beloy, Shimon Kolkowitz, Ronen Kroeze, Travis Nicholson,
Thibault Peyronel, Alp Sipahigil, Jeff Thompson, and Leo Zhou for enlightening
discussions.
This work was supported by NSF, CUA, NIST, NASA, Simons Foundation, AFOSR MURI,
ARL and NSSEFF fellowship.

\begin{appendix}
\section*{SUPPLEMENTARY MATERIALS}
\section{Using the messenger atom}
With
proper optical control, we can entangle the ensembles by moving a single
Rydberg atom to the vicinity of each ensemble sequentially, such that its
blockade radius covers one of the clouds entirely.
Starting from the state
\bel
	\ket{g}^{
	nM}\big(\ket{s} + \ket{r_2}\big),
\eel
where the first $nM$ ket stand for the state of all atoms in the $M$
ensembles (each having $n$ atoms), and the last one represents the state of the
messenger atom. In a sequence, we imagine the messenger atom to be brought to
the vicinity of each ensemble, and the pulse sequence $[\pi/\sqrt{N}]_{g,r1},
[pi]_{f,r1}$, creates an $f$ excitation, conditioned on the state of the
messenger atom. This plays out as follows
\bal
&&\rightarrow\ket{g}^{n(M-1)}\Big(\ket{1_f}_1\ket{s} + \ket{0}_1\ket{r_2}\Big)
\\
&&\rightarrow\ket{g}^{n(M-2)}\Big(\ket{1_f}_1\ket{1_f}_2\ket{s} +
\ket{0}_1\ket{0}_2\ket{r_2}\Big)
\\
&&\vdots
\\
&&\rightarrow \Big(\ket{1_f}^{M}\ket{s} + \ket{0}^M\ket{r_2}\Big),
\eal
which then only requires the messenger atom to be measured in the $\ket{\pm} =
(\ket{s} \pm \ket{r_2})$ basis, resulting in
\bel
	\rightarrow \ket{1_f}^{M} \pm \ket{0}^{M},
\eel
the required entangled state before the final GHZ extension step.

Disregarding the technical difficulties of trapping multiple atomic ensembles in
the same vacuum chamber, this entangling method has a higher fidelity than the
previous, photon-based, protocol, since it does not suffer from the errors
affecting the photon emission, propagation and detection. We model the
imperfections of this scheme by summing the error terms $\eps_1 + \eps_2 +
\eps_3$ only (from Eq.
(\ref{eq:f1}), (\ref{eq:f2}), (\ref{eq:f3})).

\section{Overview of optimization}
In section \ref{sec:clock_precision}, we show that the figure of merit, the
precision gain with respect to non-entangled schemes, can be written as
\bel
\label{eq:G_first}
	G(N,E) = \frac{\pi}{8}e^{-E N}\sqrt{\frac{N}{\log N}},
\eel
where $N$ is the total number of entangled atoms in the global GHZ state, and
$E = E(n,\Omega)$ is the total error (contrast loss) divided by the total
number of atoms. $E$ depends on the number of atoms at a single clock, $n$, and the
Rabi-frequency of the dressing field used for local entanglement
growing.

We separate out the
minimization of $E$ (through finding the optimal $n, \Omega$
parameters), and the maximization of $G$ (through finding the optimal $N$). In
other words, we find
\bel
	G_\mathrm{max} = \max_N \,G\left(N,\,\min_{n,\Omega} E\right).
\eel
This two-step procedure gives identical results to the full optimization,
\bel
	G_\mathrm{max} = \max_{N,n,\Omega} G\Big(N, E(n,\Omega)\Big),
\eel
because both the maximum of $G$ and the optimal value of $N$ are
monotonically decreasing functions of $E$, for large $N$ (as can be seen from
\refeq{eq:G_first}).
We choose the two-step procedure because it is easier to carry out and
interpret.

\section{Local entangling errors }
The initial GHZ state is never perfect due to a series of imperfections in the
implementation. Here, we analyze the main errors responsible for lowering the
initial fidelity $F_\text{local} = [1+\exp(-\eps_\text{local})]/2$ of the GHZ
state of $n$ atoms, created via the conditional dressing scheme,  described in
the main article. We assume the following errors to be independent and small,
and we approximate $\eps_\text{local}$ with the sum
of the individual errors, $\sum_j \eps_j$. We evaluate the errors for a 2D
square lattice filled in a circular region and a 3D cubic lattice filled in a
spherical region (both of radius $R$). Where there is
a difference between the two cases, we give both results.

\subsection{Imperfect blockade}
\label{sec:imperfect_blockade}
If the blockade between the levels $r_1$ and $r_2$, $\Delta_{12}$, is not large
enough, the population transfer $g \rightarrow f$ happens even if $r_2$ is
populated by a single atom. Here, we analyze the effect of this imperfection.

Each pulse $\left[\frac{\pi}{\sqrt{n-j+1}}\right]_{g,r1}$, for $j=1,2,\dots n$
excites an average population of $\sim n
\left(\frac{\Omega}{2\Delta_{12}}\right)$ to the $r_1$ Rydberg state even if it
is detuned by $\Delta_{12}$ due to the interaction with the control atom being
in $r_2$ state. There are $n$ such pulses total, resulting in the error
\bel
	\eps_1 = \frac{n^2\Omega^2}{4} \Ev{\frac{1}{\Delta_{12}^2}},
\eel
where the average is taken over every pair of atoms in the ensemble. After
calculating this average for 2D and 3D spherical ensembles with uniform density,
we obtain
\bel
\label{eq:f1}
	\eps_1 = \left(\frac{\hbar a^3\Omega}{C_{12}^{(3)}}\right)^2\times 
	\left\{
	\begin{array}{ll}
		0.02818\, n^5 &\quad \text{(2D)} \\
		0.06079\, n^4 &\quad \text{(3D)}
	\end{array}
	\right.
	,
\eel
where $C_{12}^{(3)}$ is the dipole-dipole coefficient of the interaction between
$r_1$ and $r_2$, and $a$ is the lattice constant of the square (cubic) lattice
of the 2D (3D) ensemble.

\subsection{Decaying Rydberg states}
During the pulse sequence that induce the population transfer from $g$ to $f$,
the $r_1$ level is populated by an average of $1/2$ atoms. With constant
$\Omega$ Rabi frequency, the times of the pulse $j$ is $\frac{\pi}{\Omega
\sqrt{j}}$. The total accumulated error during the pulse sequence due to decay
or dephasing of $r_1$ Rydberg state is
\bel
	\eps_2^{(1)} = \frac{\gamma_1}{2} \frac{\pi}{\Omega}2\sum_{j=1}^n
	\frac{1}{\sqrt{j}} \approx \gamma \sqrt{n} \frac{2\pi}{\Omega}
\eel
where $\gamma_1$ is the total rate of loss (environment induced decay and
dephasing) from the Rydberg level $r_1$. The additional factor of 2 appears
because both the $g\rightarrow r_1$ and $r_1\rightarrow f$ transfers need to
happen.

In the meantime, the $r_2$ level is populated by a single atom. The decay and
dephasing of $r_2$, which we assume to be happening with rate $\gamma_2$ causes
error accumulation, which we approximate as
\bel
	\eps_2^{(2)} = \gamma_2 \frac{\pi}{\Omega}2\sum_{j=1}^n\frac{1}{\sqrt{j}}
	\approx \gamma \sqrt{n}\frac{4\pi}{\Omega}.
\eel

Although the two errors affect different components of the wavefunction, we use
their sum as an upper bound of their effect:
\bel
\label{eq:f2}
	\eps_2 = \eps_2^{(1)} + \eps_2^{(2)} = 6\pi \sqrt{n}\frac{\gamma}{\Omega}.
\eel

\subsection{Imperfect self-blockade}
During the excitation of the Rydberg state $r_1$, double excitations are mostly
shifted out of resonance by $\Delta_{11}$ due to the strong van der Waals
interaction between two $r_1$ atoms.
The time average of the population in the state where one Rydberg atom is excited is
$1/2$. The collective Rabi frequency between the 1-Rydberg state and the
2-Rydberg state is $\sqrt{2(n-1)}\Omega$. This translates to an average
population of $(n-1)\left(\frac{\Omega}{2\Delta_{11}}\right)$ during a single
pulse. Since there are $n$ such pulses during the population transfer from $g$
to $f$, the total accumulated error is
\bel
	\eps_3 \approx \frac{n^2\Omega^2}{4}\Ev{\frac{1}{\Delta_{11}^2}}.
\eel
After evaluating the average over all pair in the 2D (3D) ensemble, we
obtain
\bel
\label{eq:f3}
	\eps_3 = \left(\frac{\hbar a^6\Omega}{C_{11}^{(6)}}\right)^2 \times 
	\left\{
	\begin{array}{ll}
		0.01594\, n^8 &\quad \text{(2D)}\\
		0.05544\, n^6 &\quad \text{(3D)}\\
	\end{array}
	\right.
\eel
where $C_{11}^{(6)}$ is the van der Waals coefficient of the interaction between
two $r_1$ atoms, and $a$ is the lattice constant of the square (cubic)
lattice of the 2D (3D) ensemble.

\section{Non-local entangling errors}
Our protocol requires $K-1$ links to be set up between $K$ clocks. 
We denote the fidelity of a single connection by $F_\text{non-local} =
[1+\exp(-\eps_\text{non-local})]/2$,  and we approximate $\eps_\text{non-local}$
with the  sum of individual errors $\sum_i \eps_i$, detailed below.

\subsection{Imperfect blockade}
When exciting a single collective excitations, imperfect self-blockade can
result in leakage into double excited states. The probability of this can be
exponentially reduced by applying a smooth driving pulse. E.g., in case of a
Gaussian pulse of width $\tau$, and area $\pi$, exciting the $g
\rightarrow r_1$ transition is expected to be blocked when $r_2$ is populated, 
but it succeeds with probability $P_\text{double}$,
\bel
	P_\text{double} \approx \frac{\pi^2}{4}
	\exp\left[-\frac{(\Delta_{12}\tau)^2}{2}\right],
\eel
where $\Delta_{12} = C^{(3)}_{12}/(\hbar (2R)^3)$ is the minimal energy
shift in the ensemble due to the interaction of two atoms, one in $r_1$ and one
in $r_2$.
A detailed analysis of how different pulses affect the
transition probability can be found in \cite{Conover2011}. $P_\text{double} \ll
1$ requires
\bal
\label{eq:tau}
	&&\tau \leq \frac{\sqrt{2}}{\Delta_{12}} = 
	\left\{
	\begin{array}{ll}
	2 n^{3/2} \frac{\hbar a^3}{C^{(3)}_{12}} & \mathrm{(2D)} 
	\\
	2.7\, n  \frac{\hbar a^3}{C^{(3)}_{12}} & \mathrm{(3D)}
	\end{array}
	\right.
\eal 
in order to be small compared to the other errors.

\subsection{Rydberg state decay}
The $g\rightarrow r_1$ transition is driven with a pulse of duration $\tau$,
during which the $r_2$ level has a single excitation, which decays with rate
$\gamma_2$. The resulting error contribution, after all four photon pulses have
been generated, is
\bal
\label{eq:f4}
	&&\eps_4  = 4 \gamma_2 \tau = 
	\left\{
	\begin{array}{ll}
	8 n^{3/2} \frac{\hbar a^3 \gamma_2}{C_{12}^{(3)}} & \mathrm{(2D)}
	\\
	10.8\, n \frac{\hbar a^3 \gamma_2}{C_{12}^{(3)}} & \mathrm{(3D)}
	\end{array}
	\right.
\eal
where we used the expressions for $\tau$ from \refeq{eq:tau}.

\subsection{Photon propagation and detection errors}
The pairs of photons can get lost in the fiber during propagation and the
detection process (which is limited to 50\% for time-resolving detectors,  and
25\% for non-time-resolving ones). The two-photon heralding, however, detects
both of these errors. The remaining error comes from dark-counts of the
detectors. This affects a single link with the error
\bel
\label{eq:f5}
	\eps_5 \approx 4 \gamma_\text{dark}
	T_\text{detect} = \gamma_\text{dark} \frac{20}{n \gamma_e},
\eel
where $\gamma_\text{dark}$ is the dark count rate of the detectors,
 $T_\text{detect}$ (chosen such that a properly timed detector would have a
 chance to catch $1 -e^{-5} >  99\%$ of each photon) is the ``open time'' of the
 detector, and $\gamma_{e}$ is the spontaneous emission lifetime of the
 $\ket{e}\rightarrow \ket{f}$ transitions.
 The factor of $n$ is due to the collective enhancement of the said transition, and
 the factor of 4 is because four pulses are used in each connection.

\subsection{Memory loss}
During the creation step of each link, the state $\ket{s}$ is used as memory.
On average, every link relies on one $s$ qubit. The time it takes to attempt the
creation of a link is $\sim 2L/c$, the time it takes for a light pulse to do a
round-trip between two stations.
During this time, quantum information is stored in qubit $s$, which is subject
decoherence happening at a rate $\gamma_s$.
The infidelity of the link originating from this error is
\bel
\label{eq:f6}
	\eps_6 = 4\frac{2L}{c}	\gamma_s.
\eel
State $\ket{f}$ is assumed to be a long-lived clock state, its decoherence rate
is negligible.


\subsection{Imperfect photon collection}
Collective enhancement makes the excited atom in state $\ket{e}$ decay
preferentially to $\ket{g}$, and emit a photon directly to the spatial mode
$\bk_e$, where $\bk_e$ is the spatial frequency of the collective mode $e$.
In the implementation with Yb atoms (discussed in Section
\ref{sec:Implementation}), the decay channel to $\ket{g}$ has a close
to unity branching ratio ($\zeta = 0.99$), but due to the finite size of the
ensemble, the photon collection efficiency is decreased.  The probability of not
capturing the emitted photon is
\bal
\label{eq:f7}
	\eps_7 \approx \frac{k_e^2 w^2}{3nf} =
	\left\{
	\begin{array}{ll}
		\frac{k_e^2a^2}{3\pi f}, & \text{(2D)}\\
		\frac{k_e^2a^2}{3 n^{1/3}f} \left(\frac{3}{4\pi}\right)^{2/3}, & \text{(3D)}
	\end{array}
	\right. 
\eal
where $w$ is the radius of the ensembles cross section perpendicular to $\bk_e$,
($w = a(n/\pi)^{1/2}$ for 2D, and $w = a(3n/(4\pi))^{1/3}$ for 3D.), and $k_e =
2\pi/ (1.4\,\mu\text{m})$, and $f\sim 100$ is the finess of the cavity that we
envision using.


\section{Implementation with Yb}
\label{sec:Implementation}
We imagine using the lower levels of neutral Yb for our protocol,  
$\ket{g} = \ket{6s^2({}^1\!S_0)}$, 
$\ket{f} = \ket{6s6p({}^3\!P_0)}$, 
$\ket{s} = \ket{6s6p({}^3\!P_2)}$ and 
$\ket{e} = \ket{6s6p({}^1\!P_1)}$, and 
two Rydberg levels 
$\ket{r_1} = \ket{6s\tilde n p_{m=+1}({}^1P_1)}$ and 
$\ket{r_2} = \ket{6s\tilde n s({}^3S_1)}$ 
with the same principle quantum number $\tilde n$. In the
case of the 2D lattice, we set the quantization axis perpendicular to the plane
in which the atoms reside, this way the dipole-dipole interaction between two
atoms, one in $\ket{r_2}$ and the other in $\ket{r_1}$, depends only on their
separation, $|\br_1 - \br_2|$. In the case of the 3D lattice, we rely on the
overwhelming strength of the Rydberg interaction to produce reliable blockade
even between atoms in different horizontal planes.

\subsection{Rydberg lifetimes}
We use the measured values from \cite{Fang2011} for principle quantum numbers
$\tilde n\sim 20-30$, and extrapolate the inverse lifetimes of the Rydberg
states
\bel
	\gamma_1\approx\gamma_2 = \gamma = \frac{8.403\times
	10^8~\text{s}^{-1}}{(\tilde n-4.279)^3}
\eel
where $\tilde n$ is the principle quantum number of the Rydberg orbit. Although
the measurement was carried out at $300~\text{K}$, the
contribution of the black body radiation (at $\tilde n \sim 20-30$) is
negligible even at this temperature, and therefore our extrapolation accurately
describes the effect of spontaneous emission on the lifetime. Cooling of the
radiation environment will be necessary to reach the above lifetime at $\tilde
n\sim 100$ and above. Furthermore, the photoionization rate in a trapping field
with $10^4~\text{W}/\text{cm}^2$ intensity is also more than one order of magnitude smaller.

\subsection{Self-blockade, $\Delta_{11}$}
The long-range interaction between two $r_1$ atoms at a distance $R$ is
dominated by the van der Waals potential,
\bel
\label{eq:Delta_{11}(a)} 
	\Delta_{11}(R) = \frac{C_{11}^{(6)}}{\hbar R^6},
\eel
where $C_{11}^{(6)}$ strongly depends on the principle quantum number $\tilde
n$. We use results from \cite{Topcu2015}, and extrapolate the $C_{11}^{(6)}$
coefficient to high principle quantum numbers with the following formula,
\bel
	C_{11}^{(6)} = (-0.116 + 0.0339\, \tilde n) \,\tilde n^{11}\,
	\text{a.u.}
\eel
where the a.u. stands for atomic units, $E_h a_0^6 = 9.573\times
10^{-80}~\text{Jm}^6$, where $E_h$ is the Hartree energy and $a_0$ is the Bohr
radius.

\subsection{Cross-blockade, $\Delta_{12}$}
The long-range interaction between an $r_1$ and an $r_2$ atoms at a distance $R$
is dominated by the dipole-dipole interaction. We assume that the atoms are
confined in the $xy$ plane, and  because the $6s\tilde n p_{m=+1}$ state is
polarized in the $z$ direction, the interaction strength is independent of the
relative direction of one atom to the other.
\bel
\label{eq:Delta_{12}(a)}
	\Delta_{12}(R) = \frac{C_{12}^{(3)}}{\hbar R^3},
\eel
where $C_{12}^{(3)}$ depends strongly on the principle quantum number $\tilde
n$. We use results from \cite{Topcu2015}, and extrapolate the $C_{12}^{(3)}$
coefficient to high principle quantum numbers with the following formula,
\bel
	C_{12}^{(3)} = (0.149 + 0.00077\, \tilde n) \,\tilde n^{4}\,
	\text{a.u.}
\eel
where the a.u. stands for atomic units, $E_h a_0^3 = 6.460\times
10^{-49}~\text{Jm}^3$.

\subsection{Decay rates of lower levels}
The decay rate of $\ket{s} =\ket{6s6p\,,{}^3P_2}$ is $\gamma_s = 
[14.5\,\text{s}]^{-1} = 0.069\,\text{s}^{-1}$. The decay rate of the excited
state $\ket{e} = \ket{6s6p\,,{}^1P_1}$ is $\gamma_e  = 1.8\times
10^8\,\text{s}^{-1}$.

\subsection{Photon channels}
We assume that neighboring stations are $L < 10\,\text{km}$ apart from each
other, we neglect fiber and coupling loss. We further assume that single photon
detectors have a low dark count rate, i.e.
$\gamma_\text{dark}\approx 10\,\text{s}^{-1}$.

\section{Optimization}
The total initial imperfections of a GHZ state with $N$ atoms divided into $K$
clocks, each enclosing $M$ equal-sized ensembles (each of which contain $n$
atoms) is
\bal
	\eps_\text{tot} &=& (K-1)\eps_\text{non-local}
	+ MK \eps_\text{local} 
	\\ 
	&\approx & N\left(\frac{\eps_\text{local}}{n} +
	\frac{\eps_\text{non-local}}{Mn}\right) =: NE,
\eal
where the error contributions are $\eps_\text{local} = \eps_1 + \eps_2 +
\eps_3$, and $\eps_\text{non-local} = \eps_4 + \eps_5 + \eps_6 + \eps_7$, from
Eq.
(\ref{eq:f1}, \ref{eq:f2}, \ref{eq:f3}, \ref{eq:f4}, \ref{eq:f5}, \ref{eq:f6} and
\ref{eq:f7}).

It is clear that the larger $M$ is, the smaller the error is, however $nM$ (the
number of atoms in a single clock) is limited by the current state of technology
to $(nM)_\text{opt} \sim 2500$. Independently from the total atom number, $N$,
there is an optimal ensemble size, $n_\text{opt}$, for which $E$ (the total
error per atom) is minimal.
Below we find the optimal values of the parameters $\Omega$, (the
Rabi frequency the population transfer), and $n$ (the size of the each ensemble)
for fixed values of $\tilde n$ (the principle quantum number of the Rydberg
state) and  $a = 275.75\,\text{nm}$.

Using the following dimensionless variables, $\omega = \Omega / \gamma$,
$\delta_{11} = \frac{C_{11}^{(6)}}{\hbar a^6
\gamma}$ and $\delta_{12} = \frac{C_{12}^{(3)}}{\hbar a^3\gamma}$, we can write
the error per atom as $E: = \sum_i e_i$, where the terms
are $e_i = \eps_i/n$ for $i=1,2,3$ and $e_i = \eps_i/(Mn)_\text{opt} =
\eps_i/2500$ for $i=4,5,6,7$,
\bal
	e_1 &=& \left(\frac{\omega}{\delta_{12}}\right)^2\times 
	\left\{
	\begin{array}{ll}
		0.02818\, n^4 &\quad \text{(2D)} \\
		0.06079\, n^3 &\quad \text{(3D)}
	\end{array}
	\right.
	\\
	e_2 &=& \frac{6\pi}{n^{1/2}\omega}
	\\
	e_3 &=& \left(\frac{\omega}{\delta_{11}}\right)^2 \times 
	\left\{
	\begin{array}{ll}
		0.01594\, n^7 &\quad \text{(2D)}\\
		0.05544\, n^5 &\quad \text{(3D)}\\
	\end{array}
	\right.
	\\
	e_4 &=& \frac{1}{\delta_{12}\times 2500} \times
	\left\{
	\begin{array}{ll}
	8 \,n^{3/2} & \mathrm{(2D)}
	\\
	10.8 \,n & \mathrm{(3D)}
	\end{array}
	\right.
	\\
	e_5 &=&
	7.6\times 10^{-5}\, \frac{1}{2500\times n}
	\\
	e_6 &=& 1.8\times 10^{-5}/2500
	\\
	e_7 &=&
	\frac{1.532}{3\times 10^2 \times 2500} \times
	\left\{
	\begin{array}{ll}
		\frac{1}{\pi} & \text{(2D)}\\
		\frac{1}{n^{1/3}} \left(\frac{3}{4\pi}\right)^{2/3} &
		\text{(3D)}
	\end{array}
	\right.
\eal

\subsection{Optimal parameters}
\begin{figure}[h] \centering
\includegraphics[width=0.48\textwidth]{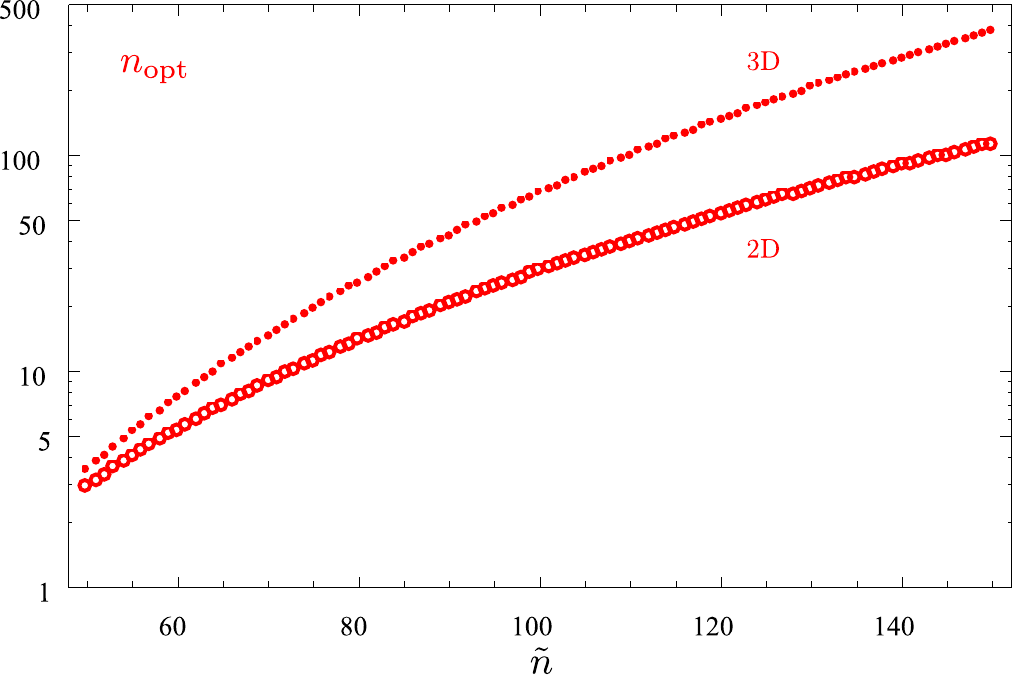}  
\caption{
\label{fig:opts}
The
optimal number of atoms in a single ensemble $n$ is plotted as a function of the
principle quantum number of the Rydberg levels $\tilde n$, for the 2D and 3D
setup.}
\end{figure}

We numerically minimized the sum, $E = \sum_i e_i$, by finding the optimal
values of $n$ for every $\tilde n \in [50,150]$, for $\omega = 10^5$. The optimal
number of atoms at a single ensemble $n_\text{opt}$ are shown on
\reffig{fig:opts}.

\begin{figure}[h]
\centering
\includegraphics[width=0.48\textwidth]{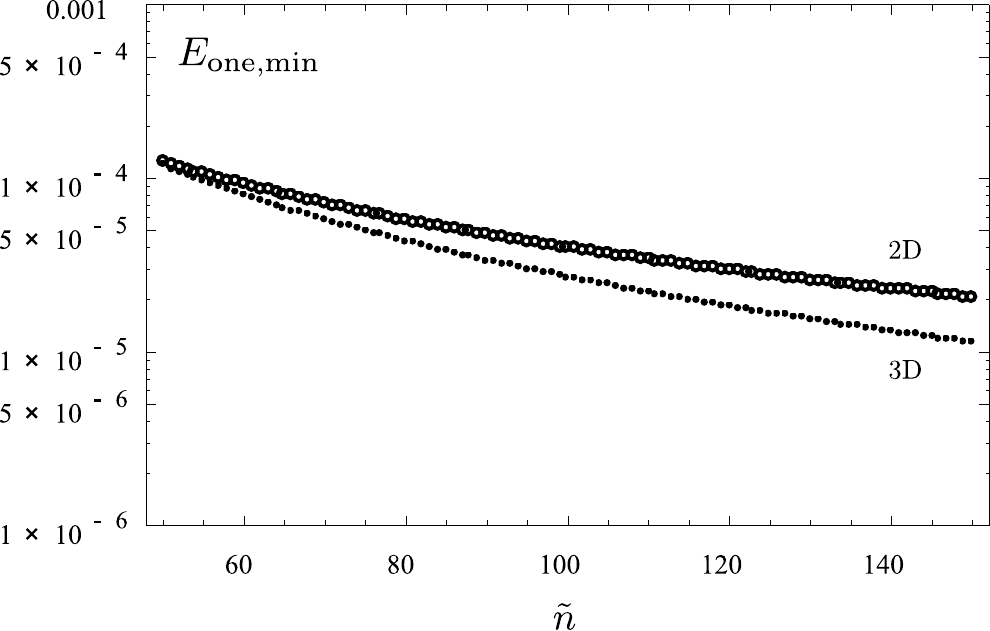}
\caption{
\label{fig:E_min}
The minimized error contribution of a single atom as a function of the principle
quantum number of the Rydberg levels $\tilde n$, for the 2D and 3D setup.}
\end{figure}

The
minimal error per atom $E_\text{min}$ is shown on
\reffig{fig:E_min}  as a function of $\tilde n$.

\subsection{Comparison of error sources}
We compare the contributions of the different error terms $e_i$ to the total
error per atom, $\sum_i e_i$, for $\tilde n = 120$.
The different error terms contribute to the sum with amounts given in Table
\ref{table:errors_2D} and \ref{table:errors_3D}.
\begin{table}
\centering
\begin{tabular}{|l|c|c|}
\hline  
 Errors in 2D ensemble & error per atom & ratio in total\\
\hline
imperfect blockade ($e_1$) 	& $3.2 \times 10^{-6}$ 	& 11\%\\
Rydberg decay ($e_2$) 		& $2.5 \times 10^{-5}$ 	& 87\%\\
self-blockade ($e_3$) 		& $\sim 10^{-10}$ 	& $<0.1$\%\\
$r_2$ decay (non-local) ($e_4$) & $ \sim 10^{-11}$ 	& $<0.1$\%\\
photon detection ($e_5$) 	& $ \sim 10^{-12}$ 		& $<0.1$\%\\
memory error ($e_6$) 		& $ \sim 10^{-9}$ 	& $<0.1$\%\\
photon collection ($e_7$) & $6.5 \times 10^{-7}$ & $2$\%\\
\hline
total error per atom 		& $3.0 \times 10^{-5}$ & 100\%\\
\hline
\end{tabular}
\caption{
\label{table:errors_2D}
The absolute and relative contribution of the different error sources to the
total error per atom at $\tilde n = 120$, $\Omega = 
10^5\,\gamma$ and $n = n_\text{opt} = 54$.}
\end{table}

\begin{table}
\centering
\begin{tabular}{|l|c|c|}
\hline
 Errors in 3D ensemble & error per atom & ratio in total\\
\hline
imperfect blockade ($e_1$) 	& $2.6 \times 10^{-6}$ 	& 14\%\\
Rydberg decay ($e_2$) 		& $1.6 \times 10^{-5}$ 	& 86\%\\
self-blockade ($e_3$) 		& $\sim 10^{-11}$ 	& $<0.1$\%\\
$r_2$ decay (non-local) ($e_4$) & $ \sim 10^{-11}$ 	& $<0.1$\%\\
photon detection ($e_5$) 	& $ \sim 10^{-12}$ 		& $<0.1$\%\\
memory error ($e_6$) 		& $ \sim 10^{-8}$ 	& $<0.1$\%\\
photon collection ($e_7$) & $\sim 10^{-8}$ & $<0.1$\%\\
\hline
total error per atom 		& $1.8 \times 10^{-5}$ 	& 100\%\\
\hline
\end{tabular}
\caption{
\label{table:errors_3D}
The absolute and relative contribution of the different error sources to the
total error per atom at $\tilde n = 120$, $\Omega = 
10^5\,\gamma$ and $n = n_\text{opt} = 146$.}
\end{table}

\section{Clock precision}
\label{sec:clock_precision}

\subsection{Imperfect initialization}
The precision of an atomic clock employing a GHZ state of $N$ clock atoms is
limited by the initial imperfect creation of the GHZ state
described by the fidelity $F_N$ or contrast $c = 2F_N -1$.
We assume that an imperfect creation of the GHZ state result in the density
matrix 
\bel
	\rho_\text{non-pure} = c \ket{\Psi}
	\bra{\Psi}
	 + \frac{1-c}{2}\big(
	\ket{\mathbf{0}}\bra{\mathbf{0}} + \ket{\mathbf{1}}\bra{\mathbf{1}} \big),
\eel
where $\ket{\Psi} = \frac{\ket{\mathbf{0}} + \ket{\mathbf{1}}
}{\sqrt{2}}$, $\ket{\mathbf{0}} = \ket{0}^{\otimes N}$, $\ket{\mathbf{1}} =
\ket{1}^{\otimes N}$, and we assumed that only the relative phase between the
two components of the GHZ state changes to an unknown value,  but no
relaxation happens.

\subsection{Measurement}
After the interrogation time, the two components of the GHZ state pick up a
relative phase $N\phi$. $\ket{\Psi}\rightarrow \ket{\Psi_\phi} =
[\ket{\mathbf{0}} + e^{iN\phi}\ket{\mathbf{1}}]/\sqrt{2}$. Performing a perfect
single-atom $-\pi/2$ rotation around the $y$ axis for all atoms transforms this
into
\bel
	\ket{\Psi'_\phi} = \frac{1}{\sqrt{2^{N+1}}}\sum_{\{q_j\}} \left[1
	+(-1)^{\sum_j q_j} e^{iN\phi}\right] \ket{q_1,q_2, \dots q_N},
\eel
where $q_j\in\{0,1\}$ stands for the state of atom $j$. After this, we 
measure every atom (in the $z$-basis). The probability of any resulting
sequence, $\mathbf{q} = (q_1, q_2, \dots q_N) \in \{0,1\}^{\times N}$, is
\bel
	\PP(\mathbf{q} | \Psi'_\phi) = \frac{1}{2^{N+1}} \left[1 + (-1)^{\sum_j
	q_j} \cos(N\phi) \right],
\eel
and the probability of the parity, $p = \big(\sum_j q_j\big)\text{ mod } 2$, is
\bel
	\PP(p|\Psi'_\phi) = \frac{1 + (-1)^p\cos(N\phi)}{2}, \qquad p\in\{0,1\}.
\eel

On the other hand, these probabilities are different when they are conditioned
on being in the mixed part of the density matrix.
\bel
	\PP(\mathbf{q}|\rho_\text{mixed}) = \frac{1}{2^N}, \qquad
	\PP(p|\rho_\text{mixed}) = \frac{1}{2}
\eel
$\forall \mathbf{q}\in\{0,1\}^{\times N}$ and $\forall p\in\{0,1\}$, where
$\rho_\text{mixed} = [\ket{\mathbf{0}}\bra{\mathbf{0}} +
\ket{\mathbf{1}}\bra{\mathbf{1}}]/2$.

The resulting total probability is the weighted sum of the
two cases,
\bal
	\PP(p|\phi) &=& c\PP(p|\Psi'_\phi) + (1-c)\PP(p|\rho_\text{mixed})
	\quad \\
	&=&
	\frac{1 + c(-1)^p\cos(N\phi)}{2},
\eal
where $c = 2F_N -1$ is the
contrast of the interference fringes.

\subsection{Fisher information}
We rely on inferring the unknown phase $\phi$, from a series of parity
measurements, as described above. The information content  (about $\phi$) of a
single measured value $p$ is quantified by the Fisher information,
\bal
	\mathcal{F}(\phi) &=& \sum_{p\in\{0,1\}} \PP(p|\phi)
	\left[\ln\frac{d}{d\phi} \PP(p|\phi)\right]^2
	\\
	&=& N^2 \frac{\sin^2(N\phi)}{1/c^2 - \cos^2(N\phi)},
\eal
where the true value of the phase is $\phi$. The
average Fisher information is
\bel
	\overline{\mathcal{F}} = \frac{1}{2\pi}\intop_{-\pi}^{+\pi} \d{\phi}
	\mathcal{F}(\phi),
\eel
which we can evaluate in the limit of $c \ll 1$,
\bel
	\overline{\mathcal{F}} \approx \frac{1}{2\pi}\intop\d{\phi} c^2 \cos^2(N\phi) =
	\frac{N^2 c^2}{2}.
\eel
In the other limit, when $1 - c \ll 1$, $F(\phi)$ is approximately $c^2$
everywhere, except near the points where $\sin(N\phi)  = 0$. We approximate the
dip at $\phi = 0$ with
\bel
	\frac{\sin^2 x}{1/c^2 - \cos^2 x} \approx \frac{x^2}{\frac{1-c^2}{c^2} +
	x^2},\qquad \text{where }\; x = N\phi,
\eel
and the integral with
\bal
	\frac{\overline{\mathcal{F}}}{N^2} &\approx& c^2 -
	\frac{2}{2\pi}\int_{-\pi}^{+\pi}\d{x} \left(1- \frac{x^2}{\frac{1-c^2}{c^2} +
	x^2}\right)
	\\
	&=& c^2 -
	\frac{\sqrt{1-c^2}}{c}  \approx 1- \sqrt{2(1-c)},
\eal
where we have used that $F$ is periodic with period $2\pi/N$.

\begin{figure}[h]
\centering
\includegraphics[width=0.4\textwidth]{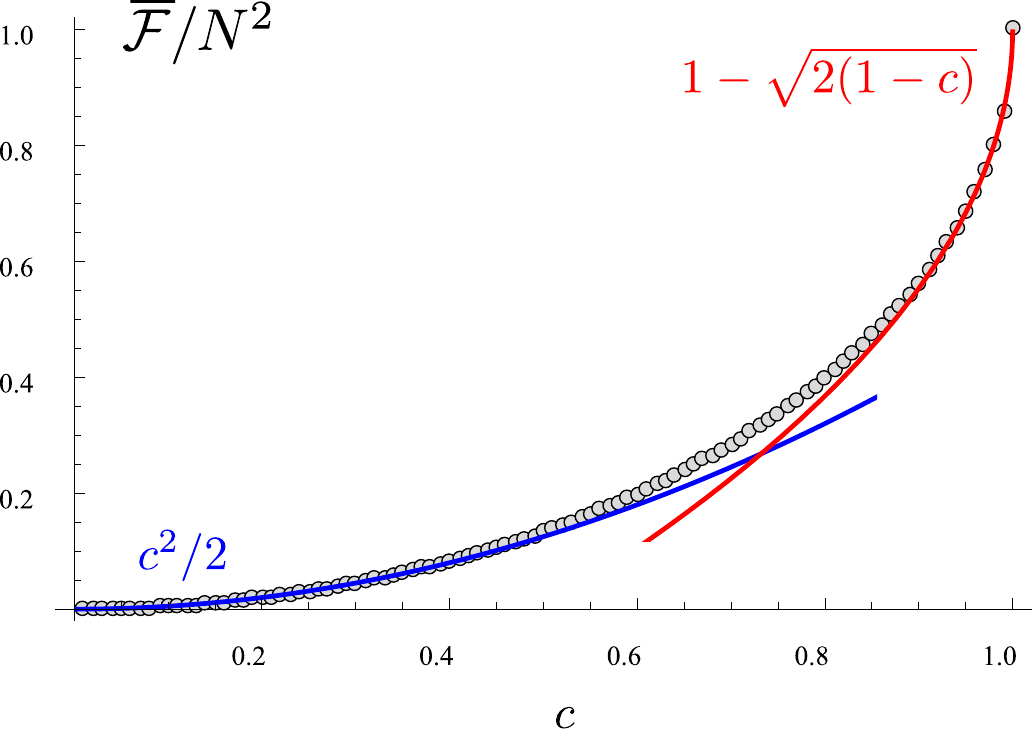}
\caption{
\label{fig:Fisher_inf}
Average Fisher information as a function of the contrast
$c$ (dots). It is well approximation by $c^2/2$ for $c < 0.6$ and by
$1-\sqrt{2(1-c)}$ for $c > 0.8$ (solid curves).} 
\end{figure}

Using these two limits for the average Fisher information, we approximate it
with
\bal
\label{eq:overline_F}
	\overline{\mathcal{F}}
	&\approx&
	\left\{
	\begin{array}{ll}
		N^2 c^2/2 &,\quad \text{if}\quad c \leq 0.7, \\
		N^2\left(1 - \sqrt{2(1-c)}\right) &,\quad \text{if}\quad 1-c > 0.7.
	\end{array} 
	\right.
\eal
The quality of this approximation can be read off from \reffig{fig:Fisher_inf}

\subsection{Cram\'{e}r-Rao bound}

The average Fisher information $\overline{\mathcal{F}}$ is a good measure of the
posterior uncertainty of the phase $\phi$, if the prior distribution of the
phase has been previously narrowed down to a small enough interval such that
its posterior is single peaked.
 In case of using the GHZ state, this requires a
very narrow prior to start with: $\phi \in [-\pi/N, +\pi/N]$. In our previous
work, we showed that this is possible by employing the atoms in a scheme using a
series of cascaded GHZ states \cite{Kessler2014}. The Cram\'{e}r-Rao bound on the
expected deviation of the estimated $\phi$ from the true one implies
\bel
	\Delta\phi = \sqrt{\Ev{(\phi_\text{estimate} - \phi_\text{true})^2} }\geq  
	\Big[\nu\overline{\mathcal{F}}\Big]^{-1/2},
\eel
where $\nu$ is the number of independent repetitions of the measurement. We are
going to assume equality to simplify our analysis.

\subsection{Allan deviation}
The average fractional frequency uncertainty of an atomic clock (with central
frequency $\omega_0$), averaged over a long time period $\tau$, is called
Allan deviation \cite{Kessler2014},
\bal
	\sigma = \frac{(\Delta\omega)_\tau}{\omega_0} \approx
	\frac{\Delta\phi_t/t}{\omega_0} \frac{1}{\sqrt{\tau/t}}
	\approx
	\frac{1}{\omega_0\sqrt{\tau}}\big[\nu t
	\overline{\mathcal{F}}\big]^{-1/2}\quad\quad
\eal
where $(\Delta \omega)_\tau =
\left|\frac{1}{\tau}\int\d{\tau'}\omega(\tau') - \omega_0\right|$ is the
deviation of the average frequency over time $\tau$, and $\Delta\phi_t$ is
the average deviation of the measured phase (from the true one) in a single
interrogation of length $t$. The $\sqrt{\tau/t}$ factor comes from the number of
independent repetitions of the same, $t$-long, interrogation cycle.

In Ref. \cite{Komar2014}, we showed that $\sigma$ can reach
\bel
	\sigma_\text{ent} \approx \frac{1}{\omega_0 \tau} \frac{8}{\pi}\frac{\sqrt{\log
	N}}{N},
\eel 
if $\tau < \gamma_\text{at}^{-1}/N$, the reduced atomic coherence time, and if
the contrast is perfect, ($c=1$). Using the approximation for
$\overline{\mathcal{F}}\approx N^2 c^2/2$,  and the fact that $\sigma \propto
[\overline{\mathcal{F}}]^{-1/2}\propto c^{-1}$, we can augment this result with
a $c$-dependence, and express the Allan deviation in the presence of
imperfections as
\bel
	\sigma_\text{ent}^\text{(imperfect)} = \sigma_\text{ent}/c = \frac{1}{c\omega_0
	\tau}\frac{8}{\pi}\frac{\sqrt{\log N}}{N}.
\eel

\subsection{Comparison to non-entangled interrogation}
Using the same number of atoms, $N$, we can arrange a measurement without using
any entanglement. This results in the Allan deviation of
\bel
\label{eq:sigma_single}
	\sigma_\text{non-ent}(\tau) \approx \frac{1}{\omega_0 \tau\sqrt{N}},\qquad
	\text{if}\quad \tau < 1/\gamma_\text{LO},
\eel
where $\gamma_\text{LO}^{-1}$ is the laser coherence time.
This, representing the standard quantum limit (SQL), is expected to be larger
than the Allan deviation corresponding to the GHZ state scheme, which is almost
at the Heisenberg limit. The precision gain of the GHZ scheme over the
non-entangled one is
\bel
\label{eq:gain_supp}
	G = \frac{\sigma_\text{non-ent}}{\sigma_\text{ent}/c} =
	(2F_N -1)\frac{\pi}{8}\sqrt{\frac{N}{\log{N}}}.
\eel
Since the fidelity $F_N$ decreases with increasing $N$, there exist an optimal
$N_\text{opt}$, for which the gain $G$ is maximal.

\subsection{Optimal clock network size}
If each clock runs with the optimal setup ($n_\text{opt}$),
then the total error per atom, $E$, is minimal, and the total fidelity can be
written as $F_N = \left[1+e^{-E_\text{min}N}\right]/2$. Plugging this into
\refeq{eq:gain_supp} gives
\bel
	G = e^{-E_\text{min} N} \frac{\pi}{8}\sqrt{\frac{N}{\log N}},
\eel 
which takes its maximum at $N = N_\text{max} \approx \frac{1}{2E_\text{min}}$,
giving $G_\text{max} \approx \frac{\pi}{8}
\left[E_\text{min}\log\left(\frac{1}{2E_\text{min}}\right)\right]^{-1/2}$.
In the meantime the number of atoms at a single clock is  $\sim 2500$.
As a result the optimal number of clocks becomes 
\bel
	K_\text{opt} \sim \frac{N_\text{max}}{2500}.
\eel

\begin{figure}[h]
\centering
\includegraphics[width=0.48\textwidth]{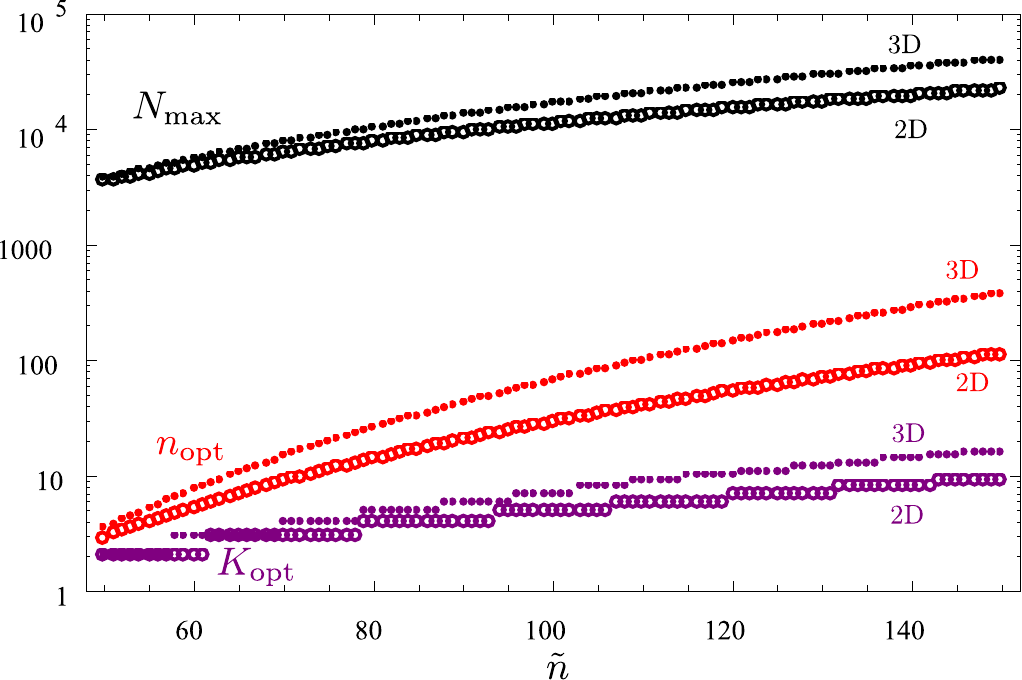}
\caption{
\label{fig:K}
The optimal total number of entangled atoms in the network $N_\text{max}$ and
the number of atoms at a single clock $n_\text{opt}$ as a function of the
principle quantum number $\tilde n$. The thin dotted lines show the multiples of
$n_\text{opt}$. The optimal number of clocks, $K_\text{opt} \sim
N_\text{max}/2500$ is written on the corresponding regions of $\tilde n$, for
the 2D and 3D setup.}
\end{figure}

On \reffig{fig:K}, we plot $N_{\text{max}}$, $n_\text{opt}$, and
$K_\text{opt}$ as a function of the principle quantum number of the
Rydberg states $\tilde n$.
For $\tilde n = 120$, we find $N_\text{max} \approx 15000$ (2D) and $\approx
25000$ (3D). Using the $n_\text{opt}$ values from before ($\approx 50$ and
$\approx 150$), we find $K_\text{opt} \sim 6$ and $\sim 10$, for 2D and 3D,
respectively. 

\begin{figure}[h] 
\centering
\includegraphics[width=0.48\textwidth]{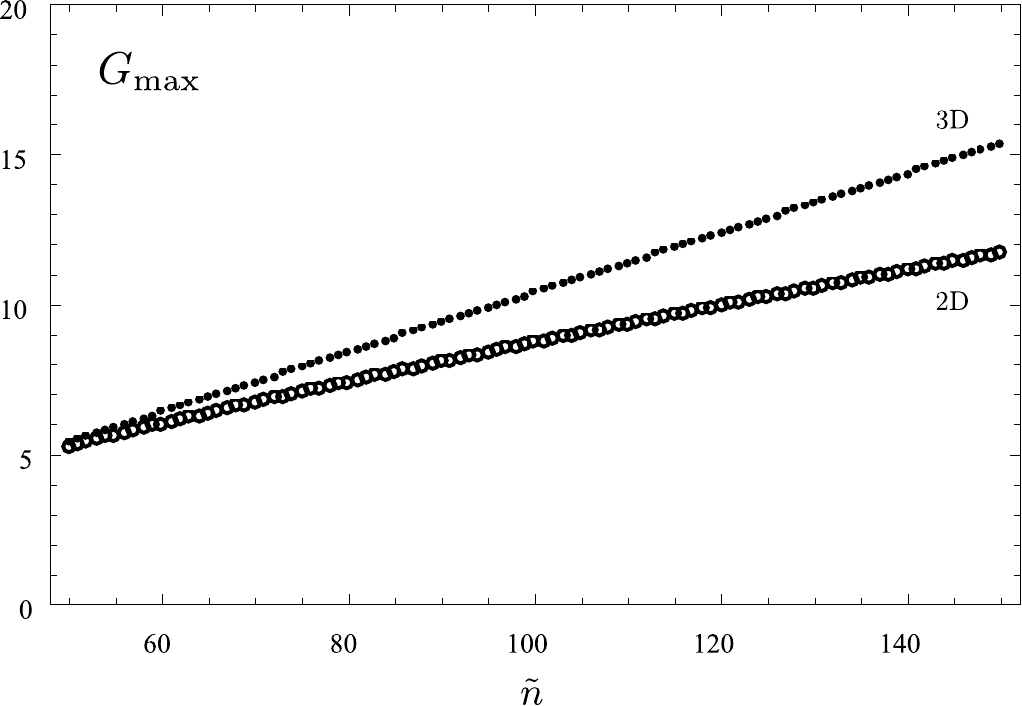}
\caption{
\label{fig:gain}
Maximal gain over the non-entangled scheme provided by the optimal entangled
clock network architecture as a function of principle quantum number of the
Rydberg states $\tilde n$, for the 2D and 3D setup.}
\end{figure}

With the optimal architecture, we can plot the maximal gain $G_\text{max}$
(compared to the non-entangled scheme using the same number of atoms) as a
function of principle quantum number $\tilde n$. 
This is shown on \reffig{fig:gain}.
For $\tilde n = 120$, the gain is $G_\text{max} = 10$ (2D) and $12$ (3D).

\section{Calculating $\ev{1/\Delta_{12}^2}$}
\label{sec:calc_Delta12}
Here, we calculate the average of 
\bel
	\frac{1}{\Delta_{12}^2} = \left(\frac{\hbar}{C_{12}^{(3)}}\right)^2 |\br_1 -
	\br_2|^6
\eel
for all $(j,k)$ pairs in
an ensemble of $n$ atoms, trapped in a (square or cubic)lattice with
periodicity $a$, uniformly filling a circular 2D (spherical 3D) region of radius
$R$.

Averaging over the cloud of atoms, can be approximated by the following integral
\bel
	\Ev{\frac{1}{\Delta_{12}^2}} \approx \left(\frac{\hbar}{C_{12}^{(3)}}\right)^2
	\underbrace{\frac{1}{V^2}\intop_V\d{^\eta \br_j}\intop_V\d{^\eta \br_k}|\br_j -
	\br_j|^6}_{R^6 I}
\eel
where $\eta = 2,3$,  $V$ is the filled region, of radius $R$, in a (2D or 3D)
lattice.

\begin{figure}[h]
\centering
\includegraphics[width=0.30\textwidth]{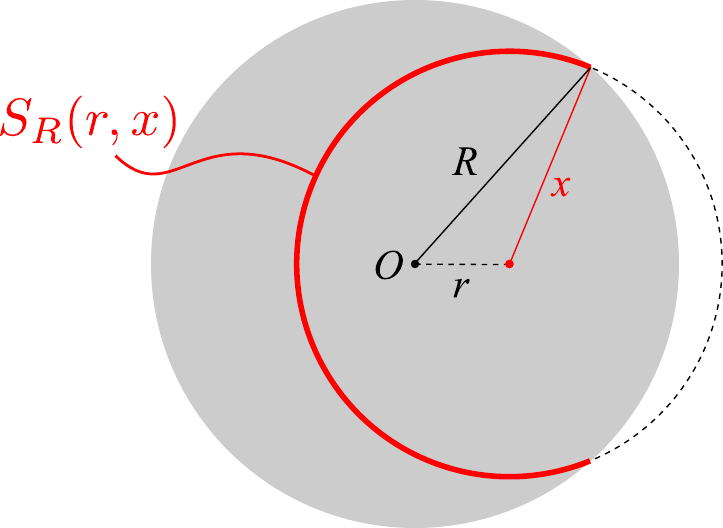}
\caption{
\label{fig:circles}
The length of the circle segment of radius $x$ lying inside the cloud of radius
$R$, $S_R(r,x)$, is between 0 and $2\pi x$ for $R-r < x < R+r$, where $r$ is
the separation between the centers.}
\end{figure}

We introduce new variables $x = |\br_j - \br_k|$, $r = |\br_j|$, and use
the circular symmetry of the cloud and the spherical symmetry of the
interaction, to turn the integrals into one dimensional ones.
\bal
	R^6 I_\text{2D} &=& \frac{1}{(\pi R^2)^2}\intop_0^R\d{r}2\pi r
	\intop_{0}^{2R}\d{x} S_R(r,x) x^6,
	\\
	R^6 I_\text{2D} &=& \frac{1}{(4\pi R^3/3)^2}\intop_0^R\d{r}4\pi r^2
	\intop_{0}^{2R}\d{x} A_R(r,x) x^6,
\eal
where the
weighting factor $S_R(r,x)$ is the length of the segment of a circle of radius
$x$, centered at $r$ distance from the origin that lies inside the 2D cloud of
radius $R$. (See \reffig{fig:circles}).
It can be written as
\bal
\label{eq:S_R}
	S_R(r,x) &=&
	\left\{
	\begin{array}{ll}
		2\pi x &,\, \text{if}\; x < R-r \\ 
		0 &,\, \text{if}\; R+r < x \\
		2x \arccos\left(\frac{x^2 + r^2 - R^2}{2xr}\right) &,\, \text{otherwise}
	\end{array}
	\right.\quad
\eal
Similarly, $A_R(r,x)$ is the area of a spherical surface or radius
$x$ centered  $r$ distance from the center of the 3D
cloud located inside the cloud.
It can be written as
\bal
\label{eq:A_R}
	A_R(r,x) &=&
	\left\{
	\begin{array}{ll}
		4\pi x^2 &,\, \mathrm{if}\; x < R-r \\
		0 &, \, \mathrm{if}\; R+r < x\\
		\pi\frac{x}{r}\left[R^2 - (x-r)^2\right] &,\, \mathrm{otherwise}
	\end{array}
	\right.\quad
\eal
Using the explicit expressions of \refeq{eq:S_R} and (\ref{eq:A_R}), we can
write
\bal
	I_\text{2D} &=& 4\intop_0^1\d{\rho}\rho \intop_0^{1-\rho}\d{\xi} \xi^7 + \\
	&&
	+ 4 \intop_0^1 \d{\rho}\rho \intop_{1-\rho}^{1+\rho}\d{\xi} \frac{1}{\pi}\xi^7
	\arccos\left(\frac{\xi^2+\rho^2 -1}{2\rho\xi}\right),
	\\
	I_\text{3D} &=& 9\intop_0^1\d{\rho}\rho^2\intop_{0}^{1-\rho}\d{\xi} \xi^8 + \\
	&&
	+ 9 \intop_0^1\d{\rho}\rho^2\intop_{1-\rho}^{1+\rho}\d{\xi}
	\frac{1}{4}\frac{\xi^7}{\rho}\left[1 - \left(\xi-\rho\right)^2\right],
\eal
which we numerically evaluate and find $I_\text{2D} = 3.5$, and
$I_\text{3D} = 4.27$.

Using that $\pi R^2 = n a^2$ (in 2D) and $4\pi R^3/3 = n a^3$ (in 3D), we can
obtain the expressions in \refeq{eq:f1}.

\section{Calculating $\ev{1/\Delta_{11}^2}$}
Following the same line of thoughts as in the previous section, we can write the
average as
\bel
	\Ev{\frac{1}{\Delta_{11}^2}} = \approx
	\left(\frac{\hbar}{C_{11}^{(6)}}\right)^2
	\underbrace{\frac{1}{V^2}\intop_V\d{^\eta \br_j}\intop_V\d{^\eta \br_k}|\br_j
	- \br_j|^{12}}_{R^{12} J}
	\frac{1}{V^2}.
\eel
The integral $J$ can be evaluated following the same methods as in the previous
section, and we obtain $J_\text{2D} = 61.29$, $J_\text{3D} = 68.26$.

Using that $\pi R^2 = n a^2$ (in 2D) and $4\pi R^3/3 = n a^3$ (in 3D), we can
obtain the expressions in \refeq{eq:f3}.

\end{appendix}

\end{document}